\documentclass[twocolumn,showpacs,preprintnumbers,amsmath,amssymb,nofootinbib,groupedaddress,superscriptaddress,prl]{revtex4}

\usepackage{docs}
\usepackage{graphicx}
\usepackage{dcolumn}
\usepackage{epsfig}
\usepackage[dvips]{color}
\usepackage{bm}
\usepackage{fancybox}
\usepackage{mathrsfs}
\usepackage{booktabs}

\allowdisplaybreaks[4]

\begin{document}

\title{%
Search for Lepton Number Violating Charged Current Processes 
with Neutrino Beams
}

\author{Shinya Kanemura}
\email{kanemu@sci.u-toyama.ac.jp}
\affiliation{%
Department of Physics, University of Toyama,
3190 Gofuku, Toyama 930-8555, Japan
}

\author{Yoshitaka Kuno}
\email{kuno@phys.sci.osaka-u.ac.jp}
\affiliation{%
Department of Physics, Osaka University, 
Toyonaka, Osaka 560-0043, Japan
}

\author{Toshihiko Ota}
\email{toshi@mppmu.mpg.de}
\affiliation{%
Max-Planck-Institut f\"{u}r Physik
(Werner-Heisenberg-Institut),
F\"{o}hringer Ring 6,
80805 M\"{u}nchen, Germany
}



%

\begin{abstract}
We propose a new idea to 
test a class of loop-induced neutrino mass mechanisms
by searching for lepton number violating charged current processes 
with incident of a neutrino beam. The expected rates of these processes 
are estimated based on some theoretical assumptions. 
They turn out to be sizable so that detection of 
such processes could be possible at near detectors 
in future highly intense neutrino-beam facilities.
\end{abstract}

\preprint{UT-HET 068, MPP-2012-89}
\maketitle

It has been well established that neutrinos are massive, by various experimental observations of neutrino oscillations in the past decade.
However, it remains unknown which type the neutrino
mass is, either a Dirac or a Majorana type.
Future experiments searching for neutrinoless double beta decays would
give us a clue.
However, even when the neutrinos turn out to be of the Majorana
type, the mechanism for the neutrino mass generation should be further determined.
There are two kinds of mechanisms to generate the Majorana neutrino mass.
One of them is a tree-level mechanism~\cite{Minkowski:1977sc,Yanagida:1979as,GellMann:1980vs,Mohapatra:1979ia,Schechter:1980gr,Bonnet:2009ej},
such as the seesaw models,
and the other is a loop-level
mechanism~\cite{Zee:1980ai,Ma:1998dn,Bonnet:2012kz}, such as the Zee model.
Since the Majorana neutrino mass violates the lepton number,
the Lepton Number Violating (LNV) interaction is a general feature of the all mechanism to generate the Majorana neutrino mass.
Therefore, in order to obtain a hint for the neutrino mass generation,
the LNV interaction should be studied.
For example, there have already been some studies on LNV
processes associated with charged leptons at colliders 
in some specific models~\cite{Keung:1983uu,Blaksley:2011ey,Aoki:2010tf,Krauss:2011ur}.
Since the LNV processes caused by the seesaw mechanism highly suppressed by small neutrino masses~\cite{Kersten:2007vk,Ibarra:2010xw}, it is very difficult to detect them experimentally.
On the other hand, it could be possible to detect the LNV interaction
caused by a class of loop-induced neutrino mass models, 
since it is not necessarily
related by the smallness of neutrino mass. 

In this letter, we discuss a new possibility of experimental detection of the LNV Charged Current (LNV-CC) interaction for the loop-induced neutrino mass mechanism.
In particular, we focus on the LNV interactions of 
the anti-symmetric combination of lepton doublets, $L$, given by
\begin{align}
\overline{L^{c}} {\rm i} \tau^{2} L S^{+},
\label{eq:Lit2LS}
\end{align}
where $S$ is a charged singlet scalar field, which is often called as the Zee singlet~\cite{Zee:1980ai}.
This contains a charged current interaction between 
a neutrino ($\nu$)  and a charged lepton ($\ell$) given by
$\overline{\nu^{c}} {\rm P}_{L} \ell S^{+}$
%
and does not have a doubly-charged current interaction between pure charged lepton combination given by
$\overline{\ell^{c}} \ell$.
Therefore, the interaction in Eq.~\eqref{eq:Lit2LS}
cannot be constrained by experimental bounds from the charged lepton
processes.
As a result, the studies of the LNV-CC processes would have large opportunity of discovery.

Experimentally, we propose the measurements of anti-lepton production by
new charged current interaction of a neutrino beam, given by
\begin{equation}
\nu + N \rightarrow \ell^{+} + X.
\label{eq:LNV-detection}
\end{equation}
Such a measurement with high statistics can be made at a neutrino near detector with a magnetic field to identify an electric charge of the charged leptons.
We also propose the measurements of LNV decays of hadrons, given by
\begin{equation}
\pi^{+} \rightarrow \mu^{+} + \overline{\nu} \quad {\rm and} \quad
{}^{A}_{Z}X \rightarrow {}_{Z+1}^{A}Y + e^{-} + \nu.
\label{eq:LNV-source}
\end{equation}
These LNV decays produce anti-neutrinos (neutrinos)
in a neutrino (an anti-neutrino) beam. 
The former in Eq.~\eqref{eq:LNV-source} corresponds to a source of
conventional super neutrino beam and the latter 
affects that of a beta neutrino beam.
Therefore, the measurement at a near detector can be used to identify these LNV decays of hadrons.
The current direct upper limits to LNV-CC interaction are not extremely tight.
According to the Particle Data Group~\cite{Nakamura:2010zzi},
the LNV-CC processes have been searched for 
in neutrino oscillation experiments, which give the bound 
$\Gamma (\pi^{+} \rightarrow \mu^{+} \bar{\nu}_{e})<1.5 \cdot 10^{-3}$.
We can expect that future neutrino oscillation experiments with high neutrino intensity 
will improve significantly this type of direct bounds on 
the LNV-CC process with high precision.

\begin{figure}[t]
\unitlength=1cm
\begin{picture}(4,3)
\put(0,0.2){\includegraphics[width=4cm]{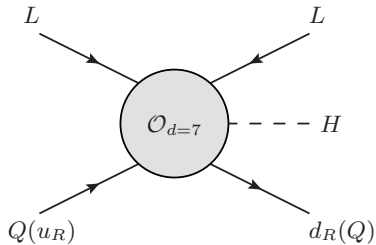}}
\put(-0.2,0){$Q(u_{R})$}
\put(3.8,0){$d_{R}(Q)$}
\put(0,2.85){$L$}
\put(3.8,2.85){$L$}
\put(3.95,1.4){$H$}
\put(1.65,1.4){$\mathcal{O}_{d=7}$}
\end{picture}
\caption{Gauge invariant realization of LNV interactions
 $\overline{\ell^{c}} \nu \overline{d} u$,
 which must include a Higgs doublet field.}
\label{Fig:CC-LNV-gaugeinv}
\end{figure}

The effective four-Fermi LNV Lagrangian for the charged current interaction with quarks can be parameterized by
\begin{align}
\mathscr{L}
=
2\sqrt{2} G_{F}
(\mathcal{C}_{L/R})_{i}^{\beta \alpha}
(\mathcal{O}_{L/R})^{i}_{\beta \alpha}
+{\rm H.c.},
\label{eq:Leff-after-EWSB}
\end{align}
where $G_{F}$ is the Fermi constant, $\mathcal{C}_{L/R}$ are 
mass dimensionless coefficients, and  $\mathcal{O}_{L/R}$ are the operators of mass dimension-six, defined as 
\begin{align}
(\mathcal{O}_{L/R})^{i}_{\beta \alpha}
\equiv
[\overline{\nu^{c}}_{\beta} {\rm P}_{L} \ell_{\alpha}]
[\overline{d}_{i} {\rm P}_{L/R} u_{i}],
\label{eq:Oeff-after-EWSB}
\end{align}
where $\alpha$, $\beta$ and $i$ are indices for flavour.
Here, for simplicity, we consider only the case where the quark flavour 
is conserved.
These effective interactions can be regarded as 
a remnant of physics at high energy scales.
After integrating out the heavy particles from the high energy models, 
the effective interactions at the electroweak scale
should be the Standard Model (SM) gauge invariant.
Therefore, the operators in Eq.(\ref{eq:Oeff-after-EWSB}) 
can be considered to be the component of the following dimension-seven 
operators with a Higgs doublet 
$H=(H^{+},H^{0})^{\sf T}$, given by
\begin{align}
(\mathcal{O}_{L})^{i}_{\beta \alpha} \subset&
[\overline{L^{c}}_{\beta} {\rm i} \tau^{2} L_{\alpha}]
[\overline{d_{R}}_{i} Q_{i} {\rm i} \tau^{2} H]  \quad {\rm and},
\label{eq:OL-realization}
\\
(\mathcal{O}_{R})^{i}_{\beta \alpha} \subset&
[\overline{L^{c}}_{\beta} {\rm i} \tau^{2} L_{\alpha}]
[\overline{Q}_{i} u_{R i} H],
\label{eq:OR-realization}
\end{align}
which are diagrammatically shown in 
Fig.~\ref{Fig:CC-LNV-gaugeinv}.
Since we focus on the anti-symmetric combination of lepton doublets
for the effective LNV-CC interaction,
the lepton flavour in the operators 
must be off-diagonal, $\beta \neq \alpha$.
Since the relevant effective operators must include 
the Higgs doublet to be kept invariant under 
the SM gauge transformation,
the LNV effect depends on the detail 
of the scalar sector in the model.
Later, 
we will specify the Yukawa sector to be one 
in the type II Two-Higgs Doublet Model (THDM).

We estimate the rates of LNV-CC signal events by a neutrino beam at a
near detector,
which can be generated by the interactions in Eq.~\eqref{eq:Leff-after-EWSB}. 
They are created in the following two different ways.
They are
(i) LNV-CC Deep Inelastic Scattering (DIS) process between neutrinos and nucleons in detectors and
(ii) anti-neutrino production by LNV-CC interaction at a neutrino beam source. 
The amplitudes of the LNV source and detection processes 
interfere with each other, although we treat them individually in this letter.
In the following, estimation of each process will be given.

Neutrinos are detected by charged leptons through the charged current interaction.
The LNV-CC interaction of Eq.~\eqref{eq:Leff-after-EWSB} would produce
charged leptons of the opposite electric charge to
the SM process, given in
\begin{equation}
\nu_{\beta} + N \rightarrow \ell^{+}_{\alpha} + X,
\label{eq:LNV-scattering}
\end{equation}
where $N$ is a target nucleus and $X$ represents all particles in the final state.
The cross-section of $\sigma_{\rm LNV}$ of the LNV-CC
DIS process in Eq.(\ref{eq:LNV-scattering}) 
is calculated 
to be
\begin{align}
\frac{{\rm d} \sigma_{\rm LNV}}{{\rm d} x {\rm d}y}
=&\sum_{i}
x \left[
f_{u_{i}}(x) + f_{\overline{d}_{i}}(x)
\right]
\left[
\left|
(\mathcal{C}_{L})_{i}^{\beta \alpha}
\right|^{2}
+
\left|
(\mathcal{C}_{R})_{i}^{\beta \alpha}
\right|^{2}
\right]
\nonumber 
\\
&\times
\frac{G_{F}^{2}s}{16 \pi} y^{2},
\label{eq:cross-section}
\end{align}
where $f_{q_{i}}$ is the Parton Distribution Function (PDF) for quark
$q_{i}$, $x$ is a longitudinal momentum fraction of parton,
$y$ is the fraction of incident neutrino energy, which is transfered
to the hadron part, and 
$s$ is the Mandelstam $s$ parameter.
Note that the LNV-CC cross-section takes the different kinematical
structure from the SM charged current process owing to its Lorenz
nature.  Therefore, it is, in
principle, possible to discriminate these signal processes from the SM processes by examining their kinematics~\cite{Kanemura:2004jt}.

\begin{figure}[t]
\unitlength=1cm
\begin{picture}(4,3.9)
\put(0,0){\includegraphics[width=4cm]{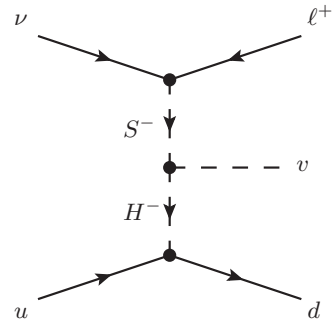}}
\put(-0.05,-0.1){$u$}
\put(3.85,-0.1){$d$}
\put(-0.05,3.8){$\nu$}
\put(3.85,3.8){$\ell^{+}$}
\put(1.4,2.3){$S^{-}$}
\put(1.4,1.2){$H^{-}$}
\put(3.7,1.85){$v$}
\end{picture}
\caption{An example of operator decomposition of the LNV 
interaction of Fig.~\ref{Fig:CC-LNV-gaugeinv},
which is motivated by the loop-induced neutrino mass model.}
\label{Fig:CC-LNV-decom}
\end{figure}
\begin{figure*}
\unitlength=1cm
\begin{picture}(18,6)
\put(0,0){\includegraphics[width=6cm]{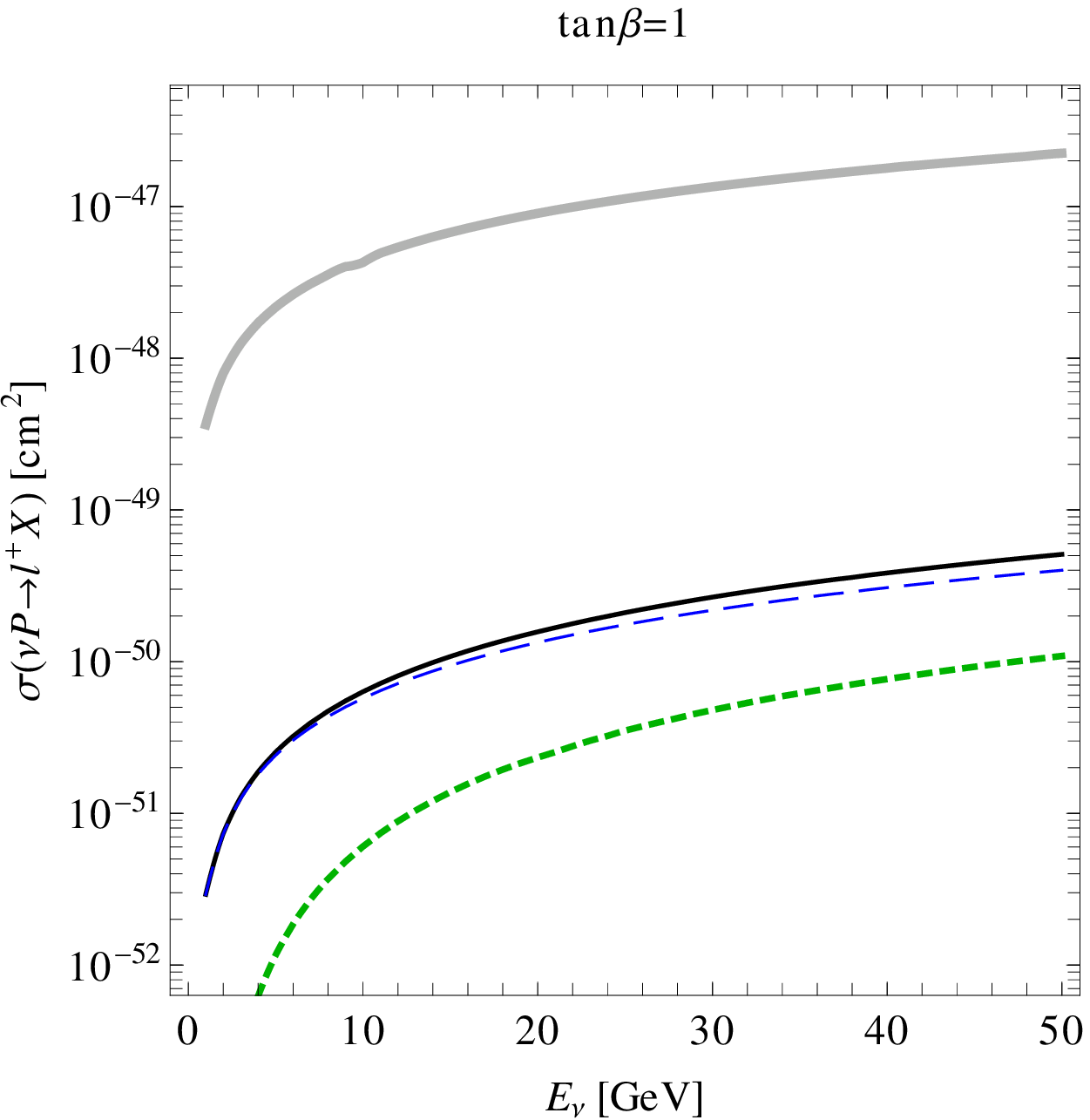}}
\put(6,0){\includegraphics[width=6cm]{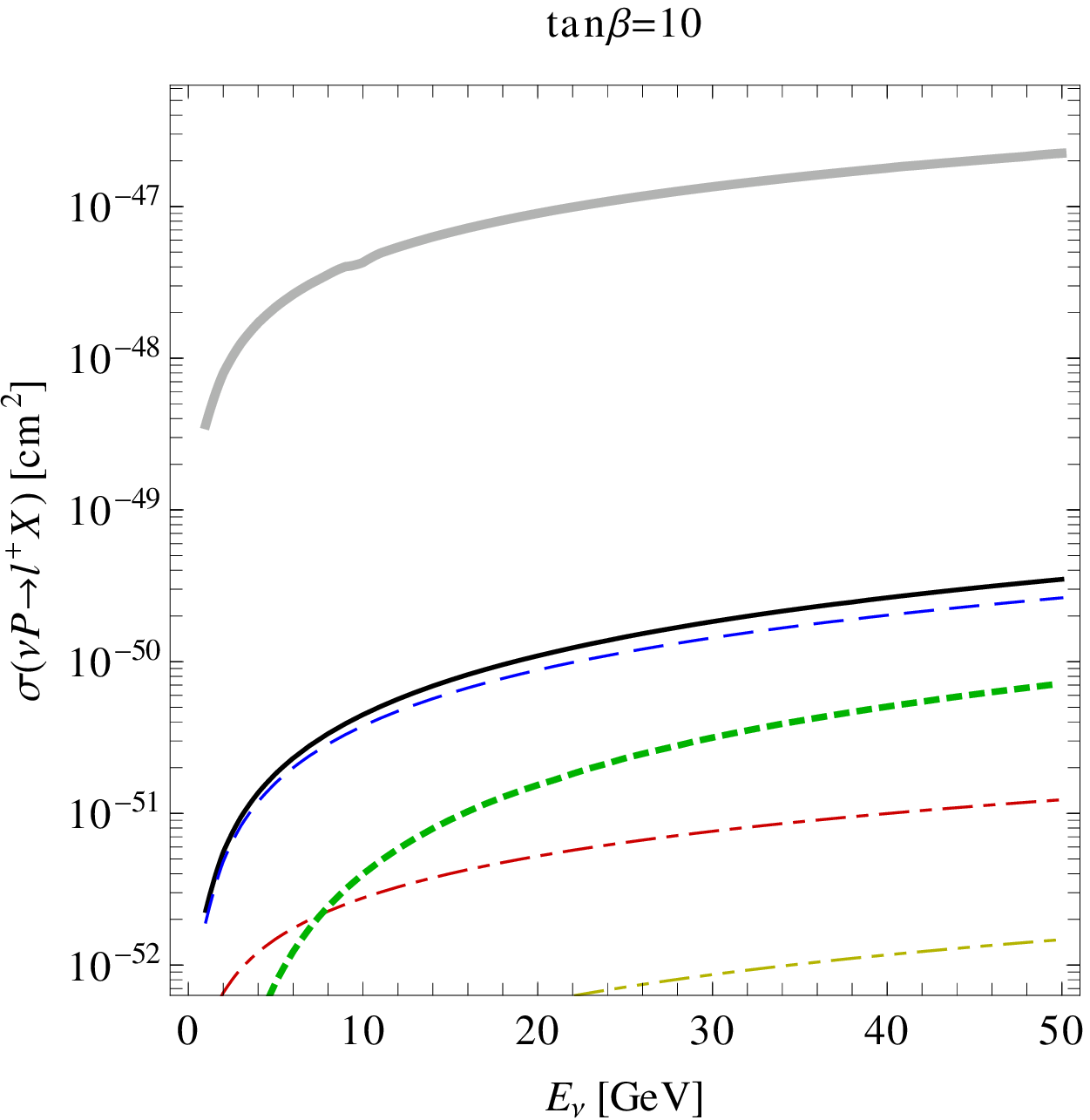}}
\put(12,0){\includegraphics[width=6cm]{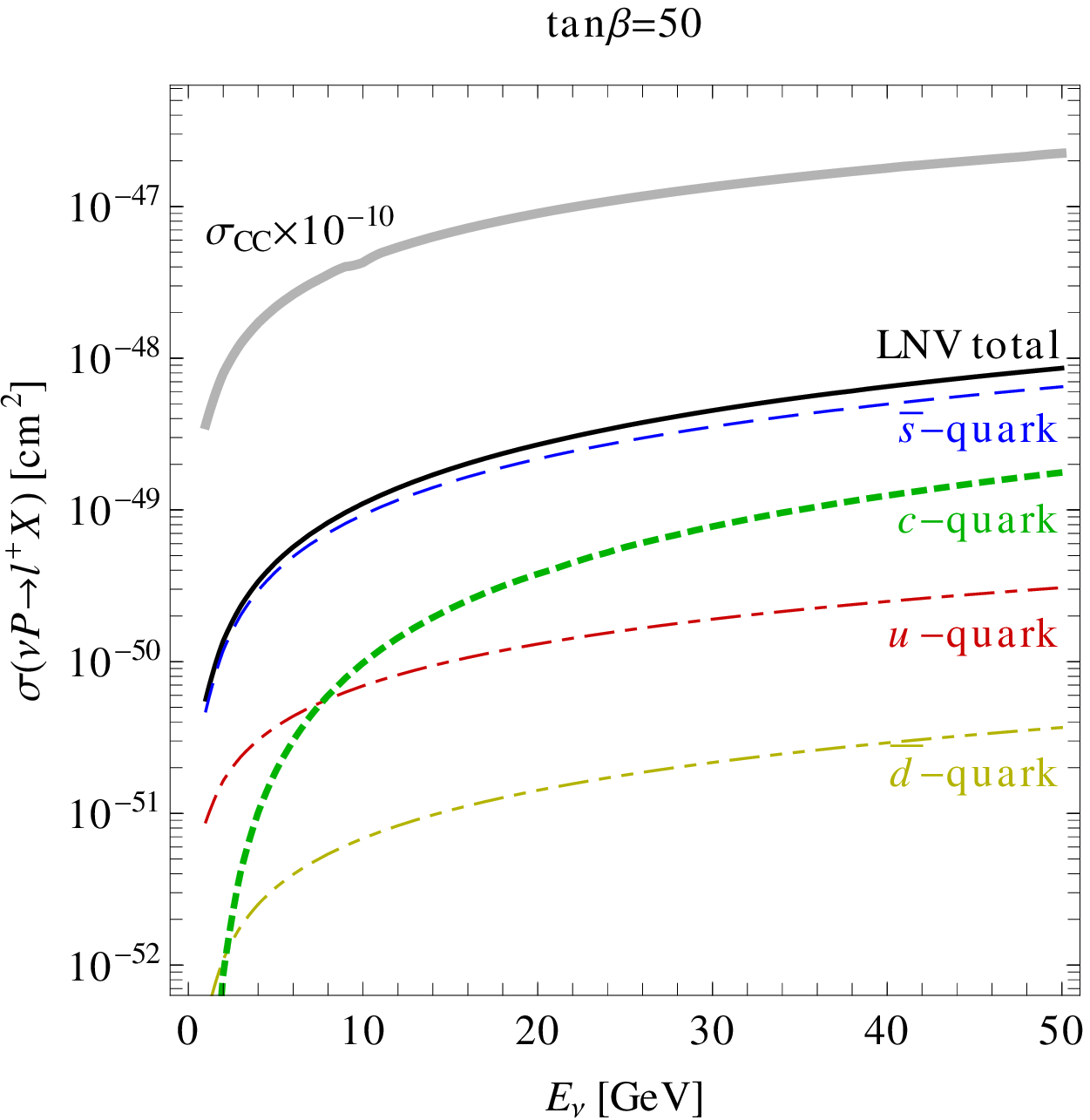}}
\end{picture}
\caption{Cross-sections of LNV-CC neutrino-proton scattering process 
(``LNV total'' solid curve)
with three different values of $\tan \beta \in \{1, 10, 50\}$.
The contributions from individual quarks are also shown
(``$q$-quarks''). 
For comparison, we plot also the cross-section (times $10^{-10}$) 
of the neutrino-proton scattering process through 
the standard model charged current (``$\sigma_{\rm CC} \times 10^{-10}$'' 
gray thick curve).}
\label{Fig:cross-section}
\end{figure*}
To estimate the signal rate, let us evaluate the magnitudes of the LNV-CC couplings  
with a set of reference values of parameters 
in typical models for loop-induced neutrino
masses.
The models are categorized into two classes
according to the source of the LNV:
(i) Lepton number is violated by Majorana nature of the heavy mediation field,
such as right-handed neutrino (e.g., Refs.~\cite{Ma:2006km,Krauss:2002px,Aoki:2008av,Kanemura:2010bq}),
and
(ii) Lepton number is transmitted to the scalar sector and explicitly violated 
by an interaction in it
(e.g., Refs.~\cite{Zee:1980ai,Babu:1988ki,Aoki:2010ib,Krauss:2002px,Kanemura:2010bq}\footnote{%
It is known that the original Zee model~\cite{Zee:1980ai} 
cannot reproduce the correct lepton mixing
matrix. To solve this problem, some extensions are necessary, 
see e.g., Ref.~\cite{Hasegawa:2003by}.}).
The latter class of the models commonly 
contains the LNV-CC interaction shown in Eq.~\eqref{eq:Lit2LS}.
We re-define it with the coupling $f^{\beta \alpha}$ as
\begin{align}
\mathscr{L}_{\rm LNV} = f^{\beta \alpha} 
\overline{L^{c}}_{\beta} {\rm i} \tau^{2} L_{\alpha} 
S^{+} + {\rm H.c.}
\label{eq:Lagrangian-LitLS}
\end{align}
Phenomenological consequences of this interaction
have been studied in various context, see e.g. 
Refs.~\cite{Cuypers:1996ia,Kanemura:2000bq,Babu:2002uu,Bergmann:2000gn,Antusch:2008tz}.
As mentioned previously, the effective four-Fermi 
LNV-CC interactions are related with the scalar sector of the models.
Here, we assume the type II THDM (e.g. Ref.~\cite{Branco:2011iw}) 
for Yukawa interactions.
The scalar sector of the models with the charged singlet field
$S^{\pm}$, 
in general, includes the interaction between $S^{\pm}$
and Higgs doublets, 
and the relevant portion of Lagrangian is presented by
\begin{align}
\mathscr{L}_{\text{scalar}} = 
\left[
\mu S^{-} H_{d}^{\dagger} H_{u}
+
{\rm H.c.}
\right]
+
M_{S}^{2} S^{+} S^{-},
\label{eq:Lagrangian-SHdHu}
\end{align}
where $\mu$ is a parameter with a unit of 
mass dimension and $M_{S}$ is the mass of $S$. 
Two Higgs doubles coupled to the $u$-type
and $d$-type quarks are given as $H_{u}$ and $H_{d}$, respectively.
The charged scalars of $S^{\pm}$ and $H^{\pm}$ 
mediate the LNV-CC interaction of 
Eq.~\eqref{eq:Leff-after-EWSB}
through the tree-level diagram shown in
Fig.~\ref{Fig:CC-LNV-decom},
where $H^\pm$ represents the physical charged
Higgs state in the THDM
in the limit of $\mu=0$.
Once a non-zero value of $\mu$ is invoked,
$S^{\pm}$ and $H^{\pm}$ are mixed (cf. e.g., Ref.~\cite{Kanemura:2000bq}). 
Within the approximation of the mass insertion of $\mu$ in the
propagation of the charged scalars 
(which are the mixture states of $S^{\pm}$ and $H^{\pm}$),
the couplings of the effective interactions 
are described with the model parameters
in Eqs.~\eqref{eq:Lagrangian-LitLS} and \eqref{eq:Lagrangian-SHdHu} 
as
\begin{align}
(\mathcal{C}_{X=[L,R]})_{i}^{\beta \alpha}
=&
\frac{f^{\beta \alpha} \mu \left[
m_{d_{i}} \tan \beta,
m_{u_{i}} \cot \beta
\right]}
{\sqrt{2} M_{S}^{2} M_{H^{\pm}}^{2} G_{F}},
%
\end{align}
where 
$\tan \beta \equiv v_{u}/v_{d}$, and $v_{u}$ and $v_{d}$ are the vacuum expectation values of $H_{u}$ and $H_{d}$, respectively. 
In this class of models, the magnitude of the LNV-CC coefficients is proportional to
the mass of the interacting quarks.
Therefore, the second generation quarks, $s$ quarks, 
in nuclei play an important role.
On the other hand, 
only the interactions with 
first generation quarks are relevant to the LNV-CC process
at a beam source (cf. Eq.~\eqref{eq:LNV-source}).
Therefore, it is suppressed by small masses of the first generation quarks.

The magnitudes of the couplings of $f^{\beta \alpha}$ 
and the mass of $M_{S}$ depend on the details of the models.
Here, we employ the values of
\begin{align}
f^{\beta \alpha} 
= 
2 \cdot 10^{-2} 
\quad
{\rm and}
\quad 
M_{S} = 600 \text{ GeV},
\end{align}
which are inspired by a model~\cite{Aoki:2010ib}
in which neutrino mass is induced at the two-loop level. 
For the parameters in the Higgs sector, we choose
\begin{align}
\mu = 200 \text{ GeV} 
\quad
{\rm and}
\quad 
M_{H^{\pm}} = 300 \text{ GeV}.
\end{align}
By using these reference values, 
the effective LNV couplings are given as
\begin{align}
(\mathcal{C}_{L})_{i=[1,2]}^{\beta \alpha}
=&
[1.9 \cdot 10^{-6}, 3.9 \cdot 10^{-5}]
\left[ \frac{\tan \beta}{50} \right],
\quad \text{and}
\label{eq:CL1-reference}
\\
(\mathcal{C}_{R})_{i=[1,2]}^{\beta \alpha}
=&
[1.9 \cdot 10^{-8},
9.7\cdot 10^{-6}] \left[ \frac{1}{\tan \beta} \right].
\label{eq:CR1-reference}
\end{align}
By substituting them into Eq.~\eqref{eq:cross-section},
the cross-section of the LNV-CC neutrino-proton 
scattering process can be estimated.
The results with different values of $\tan \beta$ 
are shown in Fig.~\ref{Fig:cross-section}.
In the numerical calculations, 
we use {\sf MSTW NNLO}~\cite{Martin:2009iq}
as PDF.
As shown in Fig.~\ref{Fig:cross-section},
the LNV-CC cross-section is dominated by 
the $s$-quark contribution.
If we work on this theoretical framework,
an extremely good charge-identification rate at near
detectors is demanded to detect the LNV-CC 
neutrino-nucleon scattering process.
Since the LNV-CC cross-section in
Eq.~\eqref{eq:cross-section} takes different kinematic structure 
from the SM one,
the angular distribution of the signal 
lepton could be used to distinguish the 
signal events~\cite{Kanemura:2004jt}. 
The total cross-section of the process in Eq.~(\ref{eq:LNV-scattering})
could be as large as $10^{-48}$ cm$^{2}$ at neutrino energy of 50 GeV. 
For this case, using $10^{20}$ neutrinos which would be available in coming experiments,
LNV-CC events of about $\mathcal{O}(100)$ can be produced  
with $\mathcal{O}(10)$ ton detector placed at $200$ m 
away from the beam front.
It is known that a neutrino beam based on pion decays (conventional
beam) has contamination of $\bar{\nu}_{e}$ and $\bar{\nu}_{\mu}$ at a level of a few \%. 
Therefore, to study LNV-CC interaction in a high precision, the
measurement of $\bar{\nu}_{\tau}$ should be attempted. In a 
neutrino beam based on decay of radioactive ions (beta beam), any kind
of anti-neutrinos can be studied.

Next, we estimate the case of anti-neutrino production by the LNV-CC interaction at a neutrino source. 
The signals are detected as events of charged leptons with 
opposite electric charge to the SM process at 
a near detector, 
which are finally the same signal as the LNV-CC
process at detection. 
The conventional neutrino beam
has an advantage to
enhance the LNV-CC signal processes,
because the pion decays in 
the SM $V-A$ interaction is  
{\it suppressed} by the factor of
$\omega_{\mu} \equiv 
(m_{\pi}/m_{\mu}) ({m_{\pi}}/(m_{u} + m_{d})) 
\sim 20$ (see e.g., Ref.~\cite{Herczeg:1995kd})
in comparison with the LNV process mediated 
by the charged scalar field. 
The branching ratio of 
the LNV pion decay process, 
\begin{equation}
\pi^{+} \rightarrow \mu^{+} + \overline{\nu}_{\alpha},
\end{equation}
induced by the effective interaction 
$(\mathcal{O}_{L/R})_{\alpha \mu}^{i=1}$ 
with a muon is calculated to be
\begin{align}
{\rm Br}(\pi^{+} \rightarrow \mu^{+} \bar{\nu}_{\alpha}) = 
\omega_{\mu}^{2} 
\Bigl| 
(\mathcal{C}_{L})_{i=1}^{\alpha \mu} 
- 
(\mathcal{C}_{R})_{i=1}^{\alpha \mu}
\Bigr|^{2}.
\end{align}
The anti-neutrino production in a beta beam 
is also affected by the interaction  
$(\mathcal{O}_{L/R})^{i=1}_{\alpha e}$
with an electron, given by
\begin{equation}
{}^{A}_{Z}X \rightarrow {}^{A}_{Z+1}Y + e^{-} + \nu_{\beta}.
\end{equation}
However, there is no enhancement mechanism 
in the case of beta decays of ions.

Although the LNV decay rate of pions is increased by 
the chiral enhancement factor $\omega_{\mu}$,
it is suppressed in the type II THDM
by the small Yukawa couplings
of the first generation quarks.
With the reference values Eqs.~\eqref{eq:CL1-reference} 
and \eqref{eq:CR1-reference}, the branching ratio is calculated 
to be 
\begin{align}
{\rm Br} (\pi^{+} \rightarrow \mu^{+} \overline{\nu}_{\alpha})
= 2.1 \cdot 10^{-9},
\end{align}
which is quite small but significantly larger than 
the LNV-CC process at the detection for the same reference 
values of model parameters.
With the same setup as in the estimation of the LNV-CC 
event rate at the detector, the source LNV-CC of $\mathcal{O}(10^{3})$ events 
can be expected.

We have proposed new measurements of LNV-CC processes to
discriminate the neutrino mass generation mechanisms. The proposed
measurements are sensitive to the loop-induced neutrino
mass models.  
We have calculated 
the rates of LNV-CC interaction, which
are sizable to be detected. 
These measurements can be done in a new-generation neutrino beam facility.



\end{document}